# Semi-Classical Analysis and Design of Quantum dot Based Electrically Pumped plasmonic nanolaser

HAMED GHODSI,[1] HASSAN KAATUZIAN,[1,*] AND ELAHE RASTEGAR PASHAKI[1]

[1]*Photonics Research Laboratory, Electrical Engineering Department, Amirkabir University of Technology, Tehran 15914, Iran.*
*\* hsnkato@aut.ac.ir*

**Abstract:** Recently there is an increasing attention to electrically pumped room temperature sub-wavelength plasmon sources because of their various potential applications mainly in the integrated plasmonic field. In this paper, a GaAs/AlGaAs quantum dot based waveguide integrated plasmonic nanolaser is introduced and theoretically investigated. Using a semi-classical rate equation model, performance of our nanolaser is studied and its characteristics are presented in details. The proposed nanolaser has a tiny footprint of **0.25 µm$^2$**, room temperature operating condition and CMOS friendly process while having remarkable output performance. The new structure generates **1µW** output power with **3.5mA** injection current ( the threshold pump current is calculated to be about **2.5µA** ) in **850 nm** and has wide modulation frequency of **10 THz** in threshold pumping rate, large Purcell factor about **3460** and high output coupling ratio to the host plasmonic waveguide.

## 1. Introduction

Plasmonics is one of the most promising technologies for fabrication of integrated circuits (IC), having both tiny footprints of electronic ICs and very high bandwidth of photonic ones. [1] However, there is a critical drawback. Despite various researches for designing plasmon sources, for instance, metallic nanoshells [2], Nanocavities [3], Nanowires [4] and waveguide-based Nanolasers [5], there is no plasmon source suitable for integration into mass production plasmonic ICs. These laboratory realizations, which in some cases have optical pumping or in some other cases just operate in cryogenic conditions cannot be used in a traditional CMOS process for commercial purposes.

Plasmon sources or plasmonic nanolasers have various potential applications like next-generation high-speed integrated circuits, nanoscale and low power light sources and medical devices [2, 3, and 6] and so on. To do so, the proposed nano laser should be electrically pumped, easy to fabricate, can be monolithically integrated into CMOS fabrication process and should have noticeable performance in room temperature. Recently several pieces of research have been done on such a device [3, 4, and 5] and hopefully, the device proposed in [5] have many of the mentioned properties. Therefore, we will use the proposed structure of [5] as a basis for our new design procedure in order to achieve even better characteristics with a more compact quantum dot (QD) based device, which is integrated into a Metal/Insulator plasmonic waveguide. Our device also in a proper bias condition can be used as an internal photoemission Schottky plasmon detector [7] without any physical alteration, which will be studied in detail in our future works.

In this paper, a GaAs/AlGaAs quantum dot (QD) nanocavity plasmon laser will be demonstrated, which can be integrated into plasmonic waveguides for the realization of next-generation integrated circuits. This proposed nanolaser as sketched in Fig.1 has several advantages over the previously introduced nanolasers. For instance, it has nearly 100% coupling efficiency to the waveguide plasmonic modes because of its thin structure (photonic modes are not allowed in this resonator size) and monolithic metal layer. In addition, the proposed nanolaser structure benefits from a large beta factor that means lower threshold and a huge Purcell factor ( 3460 in 850nm ), which leads to higher gain and better laser performance. All of these positive points are available in a tiny subwavelength footprint. In the next section, the physical structure and fabrication process will be explained. In section 3, governing rules and theories are discussed and a model for performance analysis will be introduced. In section 4, we will introduce our simulation tools and then finally in section 5, results can be witnessed. We'll also have a conclusion section 6.

## 2. Physical structure and Fabrication

The physical structure of the proposed nanolaser is sketched in Fig.1 and as can be seen consists of a cubic resonator placed over a Metal-Insulator (MI) plasmonic waveguide made from Au/SiO2 layers. The resonator is made from an **$Al_{0.3}Ga_{0.7}As$** layer for supporting GaAs QDs, covered by two metallic sheets one made from gold as a medium for generation and transfer of plasmon modes at the bottom and another made from Pd/Ti/Au alloy as an electrical contact. Gain medium of this plasmonic laser is made of uniformly distributed QDs (size of each QD is 5nm×5nm×5nm) near the Gold layer (25nm above Gold layer), which results in nearly perfect energy transfer from generated carriers to plasmonic modes. GaAs QDs have a maximum gain profile in 850nm free space wavelength and the resonator should be designed to support this wavelength.

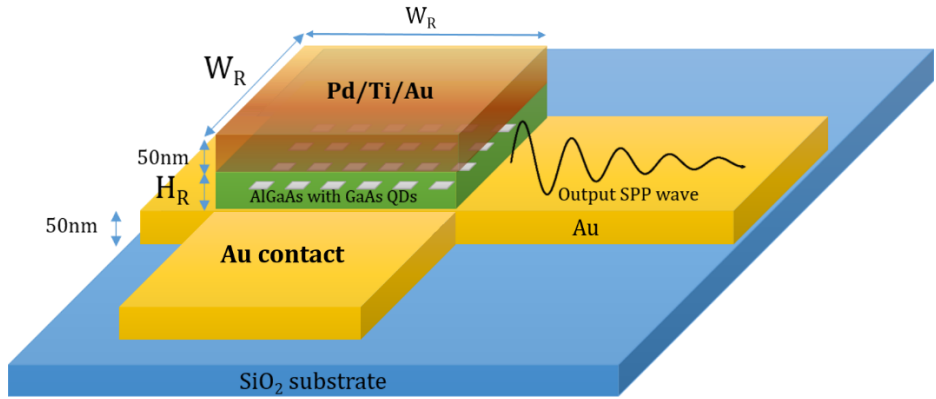

Fig. 1. 3D physical structure of the proposed device

The proposed structure will have a CMOS friendly fabrication process, starting with deposition of plasmonic waveguide metallic layer on a SiO2 substrate. Then MOCVD process is utilized to create the AlGaAs resonator as follows. First, a 25nm AlGaAs is deposited and after that, there are various techniques for growth of quantum dots as mentioned in [8] but self-assembled QDs as introduced in [9] will be an appropriate choice. Finally, there is another AlGaAs layer on the top in order to reach the total height of 50nm for the resonator. By deposition of top contact, metal (Pd/Ti/Au alloy) device fabrication will be finished.

AlGaAs and GaAs QDs are slightly p-doped (1e15cm-3) and Al alloy percent in AlGaAs layers is (30%). Dimensions of the Cubic resonator will be determined in the next section to set the resonant frequency of the cavity on the pick gain frequency of AlGaAs direct band-gap. However, size and shape of QDs and thickness of metallic layers will not be investigated in details in this paper and approximate sizes (5nm×5nm×5nm for QDs and 50nm for both metallic layers) are selected for simplicity of design. In future works, there may be more focus on the design and optimization of QDs for this structure. There can be seen a lateral 2-Dimensional schematic of layers in Fig.2.

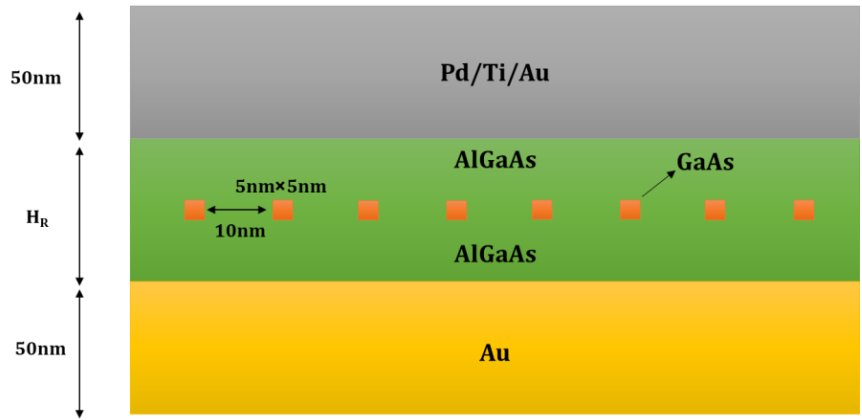

Fig. 2. 2D schematics of resonator layers

## 3. Theory of Nanolaser operation

In the proposed nanolaser structure as can be seen in Fig.3, energy of exciton generation in quantum dots due to electrical current are transferred to SPP modes at the bottom metal-semiconductor interface and the SPP flows into the host waveguide. The top metal has a considerable loss in the working frequencies and it just performs as an electrical contact.

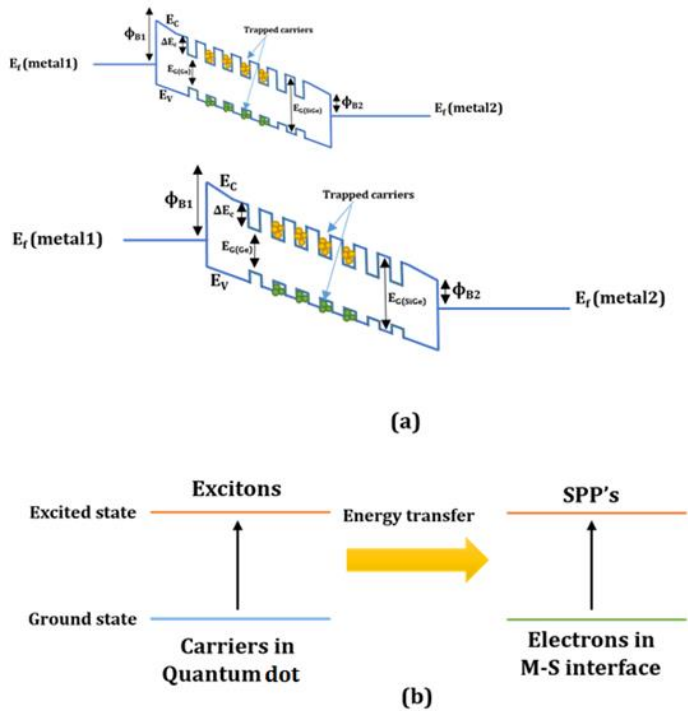

Fig. 3. Energy transfer diagram.

a. Energy band diagram of nanolaser, b. Energy transfer concept [2]

In order to analyze performance of a laser, we need a model for its rate equations. For this purpose, we'll start with semi-classic plasmonic nanolaser rate equations of (1)[10] but first, we need to discuss some concepts and to do so, we'll use two sets of theories, cavity electrodynamics, electronic carrier transport and exciton generation rate in gain medium (Quantum dots).

$$\frac{dn}{dt} = P - An - \beta \Gamma AS(n - n_0) - \frac{n v_s S_a}{V_a}$$
$$\frac{dS}{dt} = \beta An + \beta \Gamma AS(n - n_0) - \gamma S$$
(1)

In these equations "n" is the excited state population density of carriers and "S" is plasmon number in the lasing mode. "P" is the carrier generation rate in the gain medium and can be related to generation rate in quantum dots as expressed in (2):

$$P = P_1 + P_2 + ... + P_n$$
(2)

Where $P_i$'s are generation rates in the i'th quantum dot. Total carrier generation rate (P) is determined by several parameters like, pump current (Injected current by electrical pumping), thermionic emission over and tunneling through Schottky barrier, metal to QDs transit time (drift/diffusion theory), transition probability from each dot, carrier trap time in quantum dots ( indicates the average time before an exciton transfers its energy to the SPP lasing mode or lose energy due to other processes ), radiative recombination rate (specific exciton generation rate constant of GaAs), non-radiative recombination rates (Auger and SRH) and tunneling probability between two neighbor quantum dots. All of the mentioned phenomena should be considered to achieve a precise model for finding pump rate (P) as a function of pump current (A carrier dynamics model). A detailed carrier dynamic model will be introduced in our next papers because discussing the detailed models and finding a closed expression for "P" as a function of injected current, ($I_{injected}$) is not in the scope of this paper. It is worth mentioning, that the conversion efficiency in exciton/SPP energy transfer process in the effective depth of plasmonic modes is considered to be 100% as a practical approximation. [4]

"A" is the spontaneous emission rate, which can be modified by the Purcell effect via "A = $F_p A_0$", where "$A_0$" is the natural spontaneous emission rate of the material equals to $1/\tau_{sp0}$ and $\tau_{sp0}$ is the spontaneous emission lifetime of the gain medium which is GaAs QDs in our case.[18]

The Purcell factor [11] $F_p$, is a key parameter in cavity quantum electrodynamics (CQED) that defines the coupling rate between a dipolar emitter (QDs in our case) and a cavity mode. Purcell factor as can be expressed as (3) specifies the possible strategies to enhance and control light-matter interaction. [12] Efficient light-matter interaction is achieved by means of either high quality factor (Q) or low modal volume V, which is the basis of plasmonic cavity electrodynamics (PCQED).[13]

$$F_p = \frac{3}{4\pi^2}\left(\frac{\lambda}{n}\right)^3\left(\frac{Q}{V_{eff}}\right) \tag{3}$$

Where λ is free space wavelength, n is the refractive index of gain medium and Q is the quality factor of the plasmonic resonator and can be calculated by (4).[14]

$$Q = 2\pi \frac{\text{Energy stored in cavity}}{\text{Energy lost per cycle to walls}} \tag{4}$$

Which for a specific mode, it's independent of amplitude. In plasmonic metallic cavities, considering their large amount of loss, Quality factor is far less than its insulator optical counterparts are.

As mentioned before in plasmonic cavities, effective mode volume has a key role in nanolaser operation, which can be calculated by (5). [14]

$$V_{eff} = \frac{\int_V \varepsilon(r)|E(r)|^2 d^3r}{\max\left[\varepsilon(r)|E(r)|^2\right]} \tag{5}$$

Where ε is dielectric constant, E is the electric field and V is the resonator volume.

β which is known as coupling factor is defined by the ratio of the spontaneous emission rate into the lasing mode and the spontaneous emission rate into all other modes and can be expressed by (6).[5]

$$\beta = \frac{F_{cav}^{(1)}}{\sum_k F_{cav}^{(k)}} \tag{6}$$

Where $F_{cav}^{(k)}$ is the Purcell factor of k'th mode. k = 1 corresponds to the lasing mode and the summation is on both cavity modes and radiating modes. In the next section, a numerical method for calculation of coupling factor based on dipole sources will be introduced. "Γ" which equals to the ratio of carriers generated in the spatial distribution of plasmonic modes and the whole number of generated carriers, is also called mode overlap with the gain medium and will be calculated by FDTD analysis in the next section. "$n_0$" is the excited state population of carriers at transparency. $v_s$ is surface recombination velocity at the sidewalls of the resonator. "$S_a$" and "$V_a$" are the area of sidewalls of the nanolaser and volume of gain medium. Finally, "γ" is loss rate of plasmons per unit volume of the cavity (loss coefficient per unit length × modal speed/mode volume), which is calculated by $\gamma = \gamma_c + \gamma_g$. "$\gamma_c$" and "$\gamma_g$" are resonator mirror loss and loss due to the gain medium respectively [5]. Loss due to gain medium will be calculated by integrating the imaginary part of metal permittivity in the desired frequency along the path of SPPs and loss due to mirrors will be calculated by Fresnel's law in the following section.

In order to compare performance of plasmon lasers, there are several figures of merit. However, in this paper, we will use the threshold pump rate, Purcell factor, β factor, output power and operational bandwidth. Output power as can be witnessed in (7) is a function of the number of generated plasmons per unit volume of the cavity and can be derived from the rate equations of (1). [5]

$$P_{out} = \frac{1}{2} \times \frac{\alpha_m}{\alpha_m + \alpha_i} \times \frac{S}{\tau_p} \times \frac{hc}{\lambda} \times V_{mode} \tag{7}$$

Where "$\alpha_m$" and "$\alpha_i$" are mirror loss and intrinsic cavity loss respectively, "S" is plasmon number per unit volume, "$\tau_p$" is plasmon lifetime in the cavity and equals to "$Q/2\pi f_{res}$" ("Q" is the quality factor and "$f_{res}$" is the resonant frequency of the cavity), "h" is Planck constant, "c" is light speed, "$\lambda$" is the output wavelength and "$V_{mode}$" is mode volume.

The bandwidth of the proposed nanolaser is characterized by two main time constants as can be seen in (8).

$$BW = \frac{1}{2\pi}\left(\frac{1}{\tau_{elec}} + \frac{1}{\tau_{plasmon}}\right) \tag{8}$$

The first parameter is electronic delay between input switching and change in carrier generation rate "$\tau_{elec}$", which is determined by numerical carrier dynamics model introduced before. The second parameter is "$\tau_{plasmon}$" which contributes for SPP dynamics can be calculated from 3dB bandwidth of spectral response transfer function of (9). [5]

$$\tau_{plasmon} = \frac{1}{\omega_{3dB}}$$

$$\omega_{3dB} : \omega \,@\, H(\omega) = \frac{1}{2}H(0) \tag{9}$$

$$H(\omega) = \frac{\beta A(1+S_0)}{\sqrt{\left(\omega^2 - \omega_r^2\right)^2 + \omega^2 \omega_p^2}}$$

Where "$\omega_r$" and "$\omega_p$" are derived from (10) and (11) respectively and "$S_0$" is the steady-state plasmon number. [5]

$$\omega_r = \sqrt{A\left[\frac{1+\beta S_0}{\tau_p} - AN_0\beta(1-\beta)\right]} \tag{10}$$

Where "$N_0$" is steady state population inversion number. [5]

$$\omega_p = \frac{1}{\tau_p} + A(1 - \beta N_0 + \beta S_0) \tag{11}$$

## 4. Simulation methods

For utilizing the discussed theoretical models, a numerical solution framework is needed. For this purpose, the analysis is divided into two parts. At first, using FDTD method for solving the electromagnetic Maxwell equations in the cavity, parameters like quality factor, Purcell factor, mode volume and loss will be extracted and cavity dimensions are properly optimized for working in desired wavelength and bias point.

For FDTD analysis of the cavity, a rectangular three-dimensional mesh is selected and the resolution in each dimension is set to be 1 nanometer for an acceptable precision. Optical parameters of GaAs, AlGaAs, gold, contact alloy and SiO2 are extracted from experimental and theoretical models of [14] and [15] respectively. In order to extract quality factor, mode volume and Purcell factor of the cavity, resonator analysis of Lumerical FDTD solution with dipole sources are used, where the calculated optimal cavity dimension values are listed in Table.1 and effect of cavity size variations can be surveyed in Fig.4. Spatial distribution of the plasmonic modes are also simulated by MODE solution in the Lumerical software package and mode-gain medium spatial overlap parameter ($\Gamma$) is determined for the optimized cavity size of Table.1. (See Fig.5) Because of the very thin structure of the nano-laser, the distance of QDs from the Gold plate is comparable to the penetration depth of plasmonic modes and thus the "$\Gamma$" factor is calculated to be "1" which means all of QDs, interact with plasmonic modes. However, Metal plasmonic waveguide structure will provide a more important advantage in heat transfer from the resonator and the metal plate acts as an efficient heatsink for the proposed nanolaser.

**Table 1. design parameters of the cavity**

|  | Symbol | Value |
|---|---|---|
| **Cavity height** | $H_R$ | 50nm |
| **Cavity size** | $W_R$ | 250nm |
| **QD size** | $D_{QD}$ | 5nm |
| **QD separation** | $D_{QD2QD}$ | 10nm |
| **Distance of QDs from Gold plate** | $H_{QD}$ | 25nm |
| **Top/bottom metal thickness** | $H_{metal}$ | 50nm |

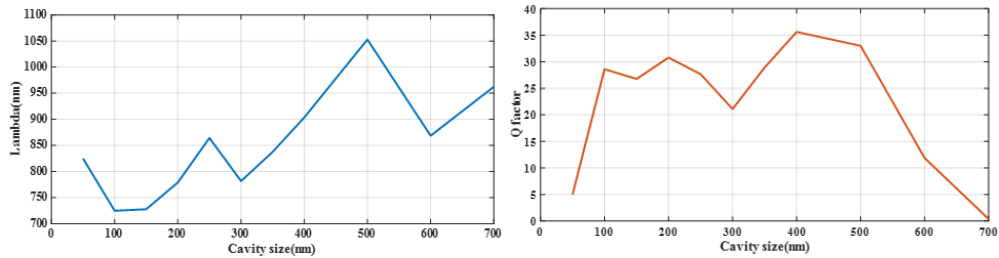

a.    b.

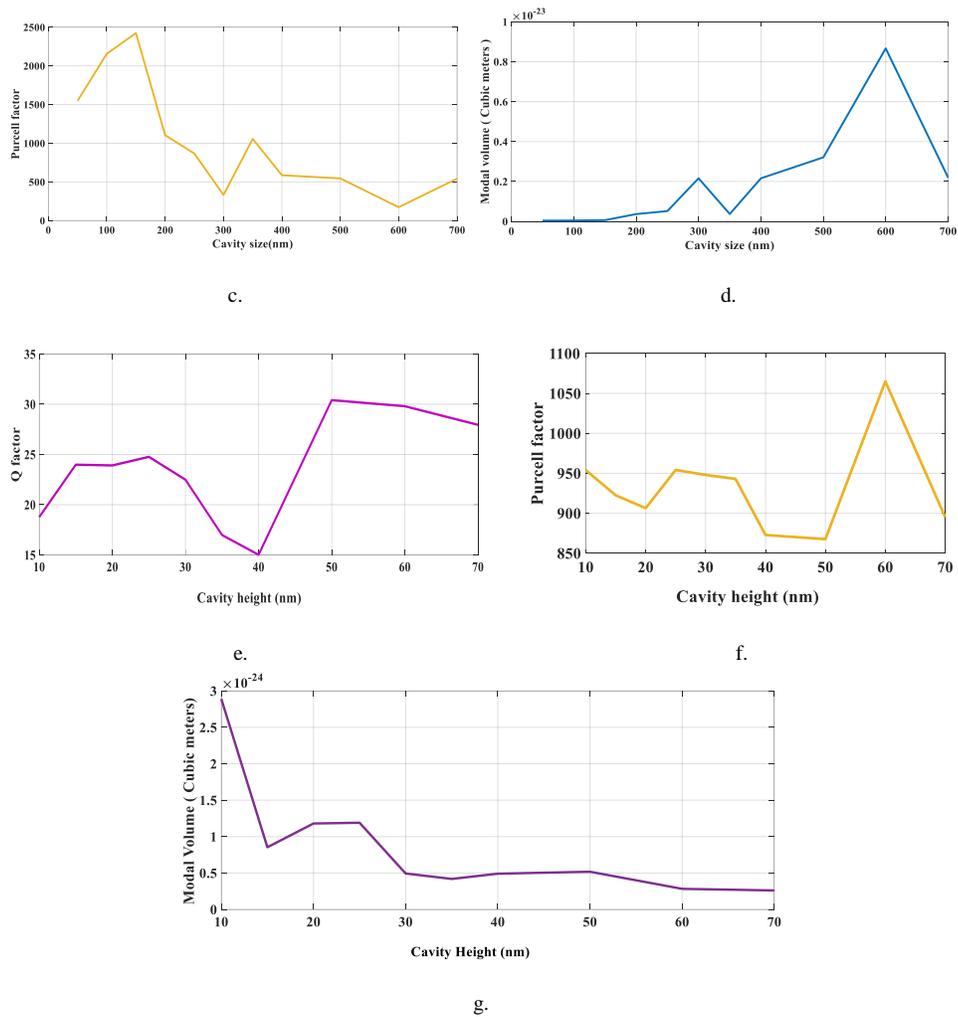

Figure4. a. λR, b. Q, c. Purcell factor and d. modal volume vs cavity size (WR) while HR = 50nm. e. Purcell factor, f. Q factor and g. modal volume vs cavity height (HR) with fixed cavity size of 250nm.

For calculating β factor, using lumerical FDTD simulation tool, a method based on several randomly positioned dipole sources is used where the lasing mode is determined by the dipole source with maximum Purcell factor. By means of (6) and calculating, Purcell factors for all of these dipole sources (As shown in figure 6), β factor is determined to be about 0.56 for the proposed square cavity structure.

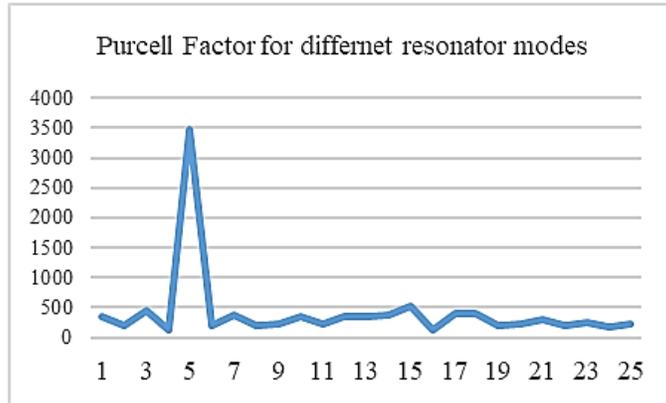

Figure.6 Change of Purcell factors of different modes by changing dipole location in the resonator

Now carrier dynamic equations should be solved for finding the exact carrier generation rates and electrical properties of the nanolaser structure. To do that, the same mesh setting as the electromagnetic FDTD code is selected and the following differential equations are solved. First universal Schottky barrier model of [16] is implemented. Then carrier transport equations of [17] and finally the quantum dot model of [18], which includes both tunneling and thermionic emission processes are implemented respectively. Furthermore, proper models for mobility, band structure, and different recombination processes are taken into account by equations of [19], [20] and [21]. The whole carrier dynamics simulation process is done by SILVACO ATLAS software package. [22]

Using the previously explained simulation process and finding the required parameters it is possible to solve the non-linear equations of (1). For this purpose and by means of a MATLAB code the rest of the equations are solved numerically and the results will be presented in the next section.

## 5. Results and Discussion

One of the most important characteristics of a laser is output power profile vs normalized pump rate ($P/P_{th}$) which is shown in Fig.7. The behavior of this profile demonstrates the proper laser operation of the introduced structure.

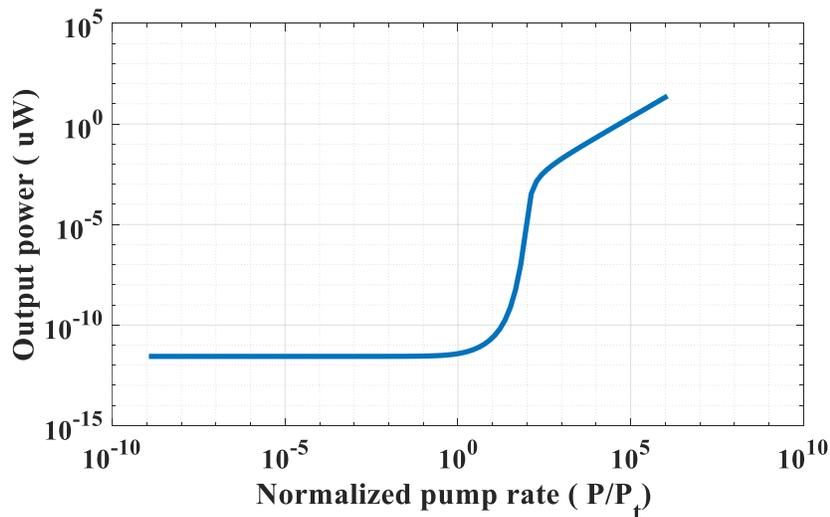

Figure.7 Output power (µW) vs normalized pump rate

In addition, in Fig.8 a better input-output characteristic that relates the output SPP power (µW) to the input injection current (µA) is shown. Relatively large output power levels while maintaining the input pump current in microampere levels and in the room temperature result in a practically appropriate device for integration processes and in order to prove this, a thermal analysis using "Lumerical Device tool" [23] was performed and thanks to the metallic waveguide structure, the temperature of the device in this relatively high pump current is lower than $80^0C$, which result in appropriate performance at room temperature without thermal breakdown.

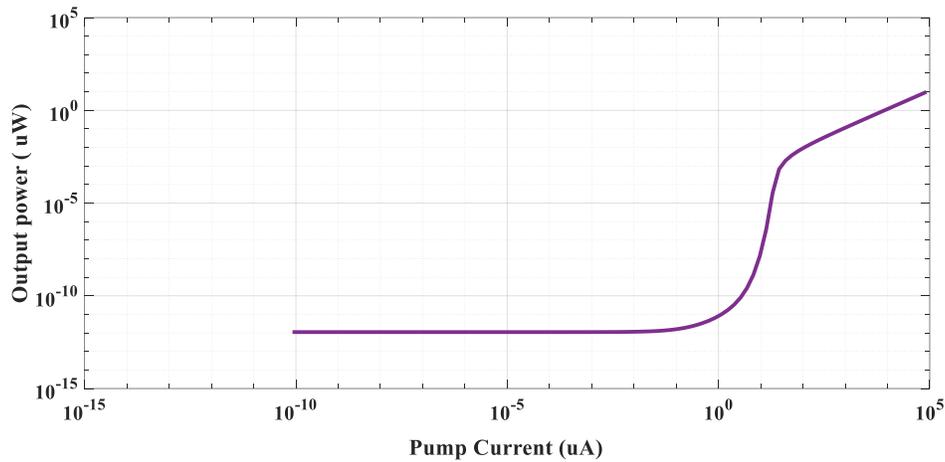

Figure.8 Output power (µW) vs injected current (µA)

Another important characteristic of the nanolaser is output power versus modulation frequency (GHz) which is depicted in Fig.9.

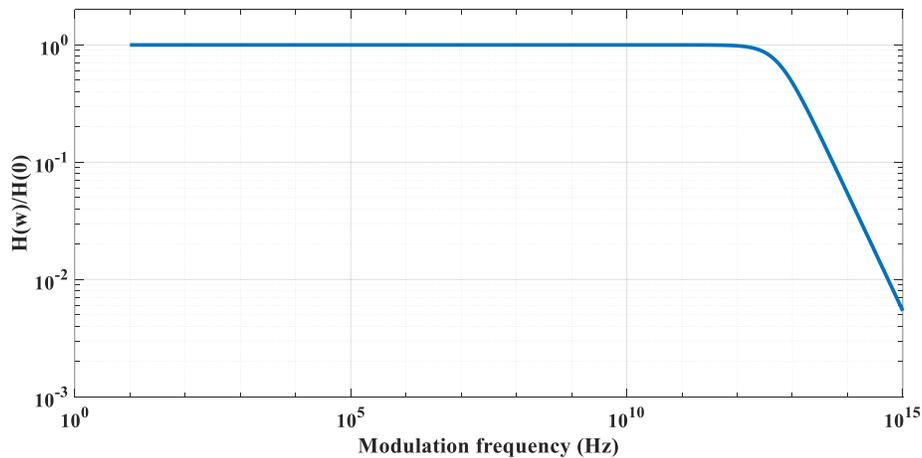

Figure.9 Output power (µW) vs modulation frequency (GHz) for $P = P_{th}$

To conclude, the key parameters of the proposed nanolaser are listed in Table.2. Although analysis done in this paper are based on theoretical models, they cannot guarantee if implemented it should work up to the derived performance. However, as mentioned before notable improvements over its competitors can be predicted.

**Table.2 Key parameters of the nanolaser**

|  | Value |
|---|---|
| **Area ( square um )** | 0.0625 |
| **Threshold current (µA)** | ~2.5 µA |
| **Output power in µW ( pump = 3.5mA )** | 1µW |
| **Modulation Bandwidth** | 10THz |
| **Purcell factor ( Lasing mode )** | 3460 |
| **Coupling factor ( β )** | 0.56 |

## 6. Conclusion

In this paper a quantum dot based waveguide integrated plasmon nanolaser was introduced, theoretically analyzed, and numerically simulated. The key advantages of the proposed structure are its tiny footprint (0.0625 µm$^2$), CMOS friendly process, room temperature operation, electrically pumping and high-efficiency coupling with metal/insulator plasmonic waveguides, which makes it a proper choice for the plasmon source in the development of plasmonic integrated circuits. The new structure generates 1µW output power with 3.5mA injection current in 850 nm, has a wide modulation frequency of 10 THz in threshold pumping rate, large Purcell factor about 3460 and very high output coupling ratio to the host plasmonic waveguide.